\documentclass[
    aps,
    prb,
    amsmath,
    amssymb,
    superscriptaddress,
    reprint
]{revtex4-2}

\usepackage{graphicx}
\usepackage{grffile}
\usepackage{siunitx}
\usepackage{tabularx}
\usepackage{csvsimple}
\usepackage[english]{babel}
\usepackage{cleveref}
\usepackage{standalone}
\usepackage{import}
\usepackage{physics}
\usepackage{bm}
\usepackage{upgreek}
\usepackage{here}
\usepackage{comment}
\usepackage{booktabs}

\usepackage{tikz}
\usetikzlibrary{
    arrows,
    arrows.meta,
	calc,
	positioning,
	optics,
	decorations,
	decorations.markings,
	decorations.text,
	arrows,
	calc}
\usepackage[version=3]{mhchem}

\begin{document}
\title{Polar and quadratic magneto-optical Kerr effects in nonmagnetic/ferromagnet bilayers for spin-orbit torque measurements}

\author{Yukihiro Marui}
\affiliation{Department of Physics, The University of Tokyo, Tokyo 113-0033, Japan}
\affiliation{Research Institute of Electrical Communication (RIEC), Tohoku University, Sendai 980-8577, Japan}

\author{Masashi Kawaguchi}
\affiliation{Department of Physics, The University of Tokyo, Tokyo 113-0033, Japan}

\author{Kohji Nakamura}
\affiliation{Graduate School of Engineering, Mie University, Tsu 514-8507, Japan}

\author{Masamitsu Hayashi}
\affiliation{Department of Physics, The University of Tokyo, Tokyo 113-0033, Japan}
\affiliation{Trans-scale quantum science institute (TSQS), The University of Tokyo, Tokyo 113-0033, Japan}

\date{\today}

\begin{abstract}
Recent studies have revealed that spin Hall magnetoresistance (SMR) contributes to both the anomalous and planar Hall resistances in nonmagnetic metal (NM)/ferromagnetic metal (FM) bilayers. This effect becomes pronounced when the NM layer exhibits a large spin Hall angle, as in W/CoFeB bilayers. In such systems, the ratio of planar to anomalous Hall resistances, normally small in single CoFeB layers, can approach unity. This unusually large ratio complicates the determination of spin-torque efficiency using harmonic Hall voltage measurements. To overcome this limitation, magneto-optical Kerr effect (MOKE) measurements have been proposed as an alternative approach.
Here, we investigate the polar and quadratic MOKE components, which correspond to, respectively, the anomalous and planar Hall resistances in the low-frequency limit to clarify whether the MOKE measurements are suitable for characterizing the spin-torque efficiency. We find that the ratio of quadratic to polar MOKE signals in NM/FM bilayers is significantly smaller than the corresponding Hall resistance ratio, indicating that SMR contributes negligibly to the MOKE response in the visible range. Consequently, the spin-torque efficiency extracted from MOKE measurements agree well with those expected from the spin Hall angle of the NM layer. These results clarify the reason why MOKE measurements provide reliable determination of the spin-torque efficiency. 
\end{abstract}

\maketitle

\section{Introduction}
Spin-orbit interaction (SOI) plays a crucial role in modern spintronics.
The large SOI of 5$d$ transition metals enables the generation of spin currents via the spin Hall effect \cite{dyakonov1971jetp,kato2004science,liu2012science}.
Spin momentum locked states emerge in systems with strong SOI and broken inversion symmetry, resulting in spin-polarized charge currents when an electric field is applied \cite{kato2004prl,miron2011nature}.
Such spin currents and current-induced spin polarizations can be exploited to manipulate the magnetization of a ferromagnetic layer.
In bilayers consisting of a nonmagnetic metal (NM) with strong SOI and a ferromagnetic metal (FM), the magnetization direction of the FM layer can be controlled by a current flowing in the film plane \cite{liu2012science,miron2011nature}.
The high efficiency and fast magnetization dynamics achievable in these systems make them attractive for technological applications.

To understand the mechanism of current-induced magnetization control, various experimental approaches have been developed.
Electrical transport measurements are widely used to evaluate the spin-torque efficiency, which characterizes the magnitude and direction of the current-induced torque acting on the magnetization.
Most such measurements exploit the anomalous Hall effect and longitudinal magnetoresistance of the FM layer to probe the magnetization direction.
These techniques can be classified into two categories: (i) those that study small changes in magnetization direction under a perturbative current and (ii) those that determine the current required to switch the magnetization.
Harmonic Hall voltage \cite{pi2010apl,kim2013nmat,garello2013nnano,hayashi2014prb,avci2015prb,roschewsky2019prb}, spin torque ferromagnetic resonance (ST-FMR) \cite{liu2011prl}, and spin Hall magnetoresistance (SMR) \cite{nakayama2013prl,kim2016prl,avci2015nphys} belong to the first class.
For the latter, the threshold current for magnetization switching is often studied as a function of external magnetic field to estimate the spin-torque efficiency \cite{yamanouchi2013apl,pai2016prb}.
Although each method has its own advantages and limitations, the spin torque efficiencies of systems such as Ta/CoFeB and Pt/Py are generally consistent across different techniques.

To explore materials with high spin-torque efficiency, it is desirable to establish a technique applicable to a wide range of systems.
However, electrical transport measurements can fail for certain materials.
The difficulty arises because the pump and probe share the same current: the current used to generate spin accumulation also serves as the probe for detecting the magnetization direction via longitudinal or Hall voltages.
When the probe voltage is influenced by spin accumulation or spin current generated by the pump, signal contamination occurs.
Such effects have been reported in harmonic Hall voltage \cite{torrejon2014ncomm,avci2015prb,yun2017npgasiamat,roschewsky2019prb}, ST-FMR \cite{karimeddiny2020prap,karimeddiny2021prap}, and SMR measurements \cite{kawaguchi2018apl,zou2016prb}.

Optical probing of magnetization via the magneto-optical Kerr effect (MOKE) \cite{zak1990jmmm,qiu2000rsi,kimel2022jpd} offers an useful alternative for studying current-induced torque \cite{fan2014ncomm,fan2016apl,chen2017prb,marui2018apex,karimeddiny2023sciadv}.
Since the pump (current) and probe (optical signal) are physically distinct, some of the limitations of electrical measurements can be avoided.
However, the principles for estimating spin-torque efficiency using MOKE is essentially the same as that in harmonic Hall voltage measurements.
That is, an alternating current is applied to induce spin current or spin accumulation in the NM layer or at the NM/FM interface, and the resulting magnetization change is detected using the Hall voltage or the MOKE signal in sync with the alternating current.
Because both signals are proportional to the off-diagonal component of the conductivity tensor, albeit at different frequencies, it is essential to understand whether optical detection can accurately quantify spin-torque efficiency in systems where electrical methods fail, and if so, to clarify the origin of the difference.

Here, we investigate the spin-torque efficiency in Ta/CoFeB and W/CoFeB bilayers, prototypical systems whose spin-torque efficiency are well studied.
We systematically compare the efficiencies obtained using electrical and optical detection schemes.
Whereas the harmonic Hall voltage measurements yield unphysically large spin torque efficiencies in W/CoFeB, consistent with previous reports \cite{torrejon2014ncomm}, the efficiencies extracted from MOKE measurements agree with those expected from the spin Hall angle of W.
Our analysis shows that the difference between the two approaches originates from the difference in the ratios of planar to anomalous Hall resistances and quadratic to polar MOKE components.
The planar-to-anomalous Hall resistance ratio approaches unity in W/CoFeB, while it remains much smaller in Ta/CoFeB.
In contrast, the quadratic-to-polar MOKE ratio is small for both bilayers.
Owing to the small quadratic MOKE component, the MOKE-based approach provides an accurate determination of the spin-torque efficiency.

\section{Model description}
\begin{figure}[ht]
    \centering
    \includegraphics[width=1.0\linewidth]{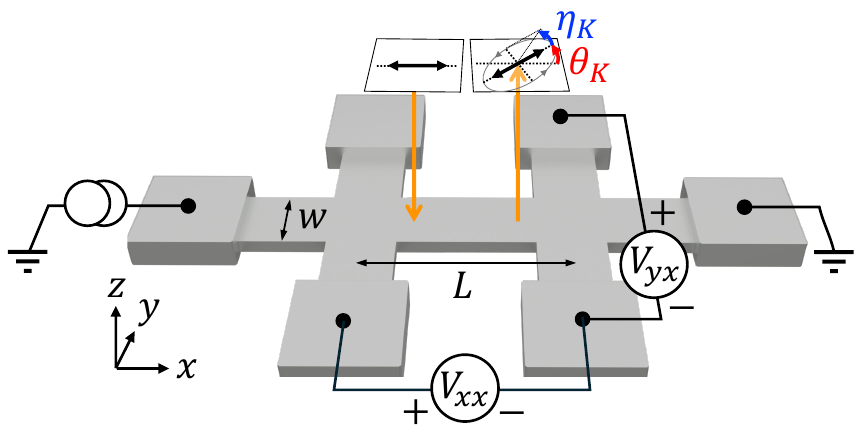}
    \caption{
Schematic illustration of the experimental setup and definitions of the coordinate axis, the longitudinal $V_{xx}$ and Hall $V_{yx}$ voltages, and the real $\theta_\mathrm{K}$ and imaginary $\eta_\mathrm{K}$ components of the MOKE signal. $w$ and $L$ are, respectively, the width and length of the current channel of the Hall bar. The polarization of the incident and reflected light are sketched. 
    }
    \label{fig:schematic}
\end{figure}

We first describe the model we use to extract the spin-torque efficiency from the harmonic Hall voltage and MOKE measurements.
We use a unit vector $\bm{m}$ to represent the magnetization direction of the ferromagnetic layer:
\begin{equation}
\label{eq:m}
    \bm{m} = (m_x, m_y, m_z) = \mqty(\sin \theta \cos \varphi, \sin \theta \sin \varphi, \cos \theta).
\end{equation}
The coordinate axis is shown in Fig.~\ref{fig:schematic}.
The film normal points along the $ z $-axis and positive current flows along $+x$.
The Hall resistance of the bilayer is given as 
\begin{equation}
\label{eq:hall}
        R_{yx} = R_\mathrm{A} m_{z} + R_\mathrm{P} m_{x} m_{y},
\end{equation}
where $R_\mathrm{A}$ and $R_\mathrm{P}$ are the anomalous Hall and planar Hall resistances.
In bulk, $R_\mathrm{A}$ and $R_\mathrm{P}$ are materials constant of the FM, but in NM/FM bilayers, they are influenced by the SMR\cite{nakayama2013prl,kim2016prl}.
In Ref.~\cite{hayashi2014prb}, $\frac{1}{2}$ was multiplied to the first term of the right-hand side.
Here, we changed the convention so that Eqs.~(\ref{eq:hall}) and (\ref{eq:kerr:mdep},\ref{eq:eta:mdep}) take the same form.
In order to reproduce the equations described in Ref.~\cite{hayashi2014prb}, one must multiply $\frac{1}{2}$ to $R_\mathrm{A}$ in the following.

The MOKE in ferromagnetic materials can be described using a similar formula.
See Fig.~\ref{fig:schematic} for a schematic illustration of the experimental setup used in this study.
The sample is irradiated with a linearly polarized light from the film normal. 
The change in the light polarization and ellipticity of the reflected light are measured.
We define a complex angle $ \Phi_\mathrm{K} $, whose real and imaginary components represent the polarization angle ($ \theta_\mathrm{K} $) and ellipticity ($ \eta_\mathrm{K} $) of the reflected light. 
\begin{equation}
\label{eq:phimoke}
    \Phi_\mathrm{K}=\theta_\mathrm{K} + i \eta_\mathrm{K}
\end{equation}
$\Phi_\mathrm{K}$ is proportional to the off-diagonal component of the conductivity tensor. Similar to the form of the Hall resistance described in Eq.~(\ref{eq:hall}), $\theta_\mathrm{K}$ and $\eta_\mathrm{K}$ can be decomposed into terms that are proportional to the out-of-plane component ($m_z$) and the products of in-plane components ($m_x m_y$) of the magnetization, provided that the light is incident from the film normal:
\begin{align}
\label{eq:kerr:mdep}
    \theta_\mathrm{K} &= \theta_\mathrm{P} m_{z} + \theta_\mathrm{Q} m_{x} m_{y},\\
\label{eq:eta:mdep}
    \eta_\mathrm{K} &= \eta_\mathrm{P} m_{z} + \eta_\mathrm{Q} m_{x} m_{y}.
\end{align}
 $\theta_\mathrm{P}$ ($\eta_\mathrm{P}$) and $\theta_\mathrm{Q}$ ($\eta_\mathrm{Q}$) are coefficients that relate $\theta_\mathrm{K}$ ($\eta_\mathrm{K}$) with the magnetization components $m_z$ and $m_x m_y$, respectively.
$\theta_\mathrm{P}$, $\eta_\mathrm{P}$ and $\theta_\mathrm{Q}$, $\eta_\mathrm{Q}$ are often referred to as the polar and quadratic (Voigt) components of MOKE, respectively.
As the responses of $\theta_\mathrm{K}$ and $\eta_\mathrm{K}$ to magnetization are essentially the same, we focus on the latter ($\eta_\mathrm{K}$) in the following.
The discussion below can also be applied to $\theta_\mathrm{K}$.

Here we assume the magnetization easy axis points along the film normal (along the $z$-axis) and define the effective magnetic anisotropy field $\bm{H}_k$ as
\begin{equation}
    \label{eq:HK}
    \bm{H}_k = (0, 0, H_k m_z),
\end{equation}
In the experiments, an external magnetic field $\bm{H}$ in the $xy$ plane is applied, that is,
\begin{equation}
\label{eq:Hext}
\bm{H} = (H_x, H_y, 0) = H (\cos \varphi_H, \sin \varphi_H, 0),
\end{equation}
where $\varphi_H$ is the angle between $\bm{H}$ and the $x$-axis.
The equilibrium magnetization $\bm{m}_0$ under $\bm{H}$ is defined as
\begin{equation}
\begin{aligned}
\label{eq:m}
    \bm{m}_0 &= \mqty(m_{0x}, m_{0y}, m_{0z})\\
     &= \mqty(\sin \theta_0 \cos \varphi_0, \sin \theta_0 \sin \varphi_0, \cos \theta_0),
\end{aligned}
\end{equation}
We assume $H$ is sufficiently small compared to $H_k$ such that the following relation holds for $\theta_0$\cite{hayashi2014prb}:
\begin{equation}
    \label{eq:theta0}
    \theta_0 \approx \frac{H}{H_k} + \mathcal{O} \left( \frac{H}{H_k} \right)^3.
\end{equation}
In addition, the magnetic anisotropy within the film plane is assumed negligible and therefore the in-plane component of $\bm{m}_0$ follows that of the external magnetic field, \textit{i.e.}, $\varphi_0 \approx \varphi_H$.
Substituting these relations into Eqs.~\eqref{eq:hall} and \eqref{eq:eta:mdep}, we obtain
\begin{align}
\label{eq:hall:thetadep}
    R_{yx} = R_{yx}^\mathrm{0} - \dfrac{R_\mathrm{A} m_\mathrm{s} - R_\mathrm{P}\sin(2\varphi_H)}{2} \left(\frac{H}{H_k} \right)^{2},\\
    \label{eq:kerr:thetadep}
    \eta_\mathrm{K} = \eta_\mathrm{K}^\mathrm{0} - \dfrac{\eta_\mathrm{P} m_\mathrm{s} - \eta_\mathrm{Q}\sin \left(2 \left(\varphi_H  - \varphi_E \right) \right)}{2} \left(\frac{H}{H_k} \right)^{2},
\end{align}
where $m_\mathrm{s} = \pm1$ characterizes the $z$ component of $\bm{m}_0$, \textit{i.e.}, $m_\mathrm{s} = \textrm{sgn}(m_{0z})$.
The function sgn$(x)$ takes the sign of $x$.
$R_{yx}^\mathrm{0}$ and $\eta_\mathrm{K}^\mathrm{0}$ are constants that depend on $m_\mathrm{s}$ but not on the magnetic field.
$\varphi_E$ is the angle between the $x$-axis and the polarization of the incident light.
Later, we will use Eqs.~(\ref{eq:hall:thetadep}) and (\ref{eq:kerr:thetadep}) to extract parameters such as $\eta_\mathrm{P}$ and $\eta_\mathrm{Q}$ [see discussion around Eq.~(\ref{eq:curvature:eta})].

Next, we describe current-induced torque measurements using the harmonic Hall voltage and MOKE measurements, following the discussion described in Ref.~\cite{hayashi2014prb}.
When a small current is passed along the $x$-axis, current-induced torque tilts the magnetization direction away from its equilibrium position $\bm{m}_0$.
Information on the current-induced torque is thus encoded in the magnetization direction, which is converted into electrical and optical signals via the Hall effect [Eq.~(\ref{eq:hall})] and MOKE [Eqs.~(\ref{eq:kerr:mdep}) and (\ref{eq:eta:mdep})].
Under an ac current with angular frequency $\omega$, the current-induced torque appears in the second harmonic signal ($2\omega$) for the Hall voltage measurements since both the resistance and current oscillate with $\omega$.
In contrast, information on the torque manifests itself in the first harmonic signal ($\omega$) for the optical measurements, which we show in the following.

The effective magnetic field associated with the torque is defined as 
\begin{equation}
    \label{eq:DH}
    \Delta \bm{H} = (\Delta H_x, \Delta H_y, \Delta H_z).
\end{equation}
In the experiments, a sinusoidal current $I$ with angular frequency $\omega$ is applied: $I = I_0 \sin (\omega t)$.
The resulting Hall voltage $V_{yx} = R_{yx} I$ is given as\cite{hayashi2014prb}
\begin{equation}
\begin{aligned}
    \label{eq:torque:Vyx}
	V_{yx} &= V_0 + V_\omega \sin (\omega t) + V_{2 \omega} \cos (2 \omega t),\\
	&V_\omega = m_s R_\mathrm{A} \left[ 1 - \left( 1 + m_s \frac{R_\mathrm{P}}{R_\mathrm{A}} \sin (2 \varphi_H) \right) \frac{H^2}{2 H_k^2} \right] I_0,\\
	&V_{2\omega} = -\frac{1}{2} \bigg[ - m_s R_\mathrm{A} ( \Delta H_x \cos\varphi_H + \Delta H_y \sin\varphi_H )\\
	& + 2 R_\mathrm{P} \big[ (- \Delta H_x \sin\varphi_H + \Delta H_y \cos\varphi_H ) \cos (2 \varphi_H) \\
	& + ( \Delta H_x \cos\varphi_H + \Delta H_y \sin\varphi_H ) \sin (2 \varphi_H) \big] \bigg] \frac{H}{H_k^2}  I_0,\\
\end{aligned}
\end{equation}
where $V_0$ is a time-independent voltage.
Note that in Eqs.~(23) and (24) of Ref.~\cite{hayashi2014prb}, terms proportional to $\sin (2 \varphi_H)$ were dropped since $\varphi_H$ was limited to 0 or $\frac{\pi}{2}$.
Here, we show all terms for completeness.
The corresponding harmonic Hall resistances are defined as:
\begin{equation}
\begin{gathered}
    \label{eq:torque:Ryx}
	R_\omega \equiv \frac{V_\omega}{I_0}, \ \ R_{2\omega} \equiv \frac{V_{2\omega}}{I_0}.
\end{gathered}
\end{equation}
Similarly, the MOKE signal change caused by current-induced torque can be derived as 
\begin{equation}
\begin{aligned}
    \label{eq:torque:eta}
	\eta_\mathrm{K} &= \eta_{0} + \eta_{\omega} \sin (\omega t),\\
	&\eta_{\omega} = \bigg[ - m_s \eta_\mathrm{P} ( \Delta H_x \cos\varphi_H + \Delta H_y \sin\varphi_H )\\
	& + \eta_\mathrm{Q} \big[ (- \Delta H_x \sin\varphi_H + \Delta H_y \cos\varphi_H ) \cos \left(2 \left(\varphi_H  - \varphi_E \right) \right) \\
	& + ( \Delta H_x \cos\varphi_H + \Delta H_y \sin\varphi_H ) \sin \left(2 \left(\varphi_H  - \varphi_E \right) \right) \big] \bigg] \frac{H}{H_k^2},
\end{aligned}
\end{equation}
where $\eta_{0}$ is a time independent MOKE signal.
As is evident, the terms inside the square bracket in Eq.~(\ref{eq:torque:eta}) take a similar form to those in $V_{2\omega}$ in Eq.~(\ref{eq:torque:Vyx}) except for the prefactor $-\frac{1}{2}$.
The factor of $-\frac{1}{2}$ in Eq.~(\ref{eq:torque:Vyx}) is introduced due to the conversion of $\sin^2 (\omega t)$ to $\cos(2 \omega t)$.
Equations~(\ref{eq:torque:Vyx}) and (\ref{eq:torque:eta}) show that one can extract information on the current-induced torque from the second harmonic Hall voltage and the first harmonic MOKE signal.

The current-induced torque is decomposed into the damping-like ($\bm{H}_\mathrm{DL}$) and field-like ($\bm{H}_\mathrm{FL}$) components.
Provided that the torque arises due to spin current and/or spin accumulation entering the ferromagnetic layer from one of its interfaces, $\bm{H}_\mathrm{DL}$ and $\bm{H}_\mathrm{FL}$ can be expressed as
\begin{equation}
\begin{aligned}
\label{eq:Hsot}
\bm{H}_\mathrm{DL} &= H_\mathrm{DL} \bm{m} \times \bm{p},\\
\bm{H}_\mathrm{FL} &= H_\mathrm{FL} \bm{p},
\end{aligned}
\end{equation}
where $\bm{p}$ dictates the axis of the polarization of spin current/accumulation.
Here, we assume the current flows along $+x$ and $\bm{p} = - \bm{e}_y$: see Ref.~\cite{lau2026arxiv} for details of the definition of $\bm{p}$.
With $\Delta H_i \equiv ( \bm{H}_\mathrm{DL} + \bm{H}_\mathrm{FL} ) \cdot \bm{e}_i$, each component reads $\Delta \bm{H} = ( m_\mathrm{s} H_\mathrm{DL}, -H_\mathrm{FL}, 0)$ where the equilibrium magnetization direction is parallel to the $z$-axis.

We define $R_{2\omega,x(y)}$ as the corresponding change in $R_{2\omega}$ when $\varphi_H = 0$ $(\frac{\pi}{2})$.
Substituting $\Delta H_x = m_\mathrm{s} H_\mathrm{DL}$ and $\Delta H_y = -H_\mathrm{FL}$ into Eq.~(\ref{eq:torque:Vyx}) and (\ref{eq:torque:Ryx}), we obtain
\begin{equation}
\begin{aligned}
\label{eq:hall:sot}
\begin{bmatrix}
R_{2\omega,x}\\
m_\mathrm{s} R_{2\omega,y}
\end{bmatrix}
&= - \frac{1}{2} \frac{H}{H_k^2} R_{\mathrm{A}}
\begin{bmatrix}
-1 &-r_\mathrm{H}\\
r_\mathrm{H} &1
\end{bmatrix}
\begin{bmatrix}
H_\mathrm{DL}\\
H_\mathrm{FL}
\end{bmatrix},\\
r_\mathrm{H} &\equiv \frac{R_{\mathrm{P}}}{R_{\mathrm{A}}}.
\end{aligned}
\end{equation}
Similarly, setting $\eta_{\omega,x(y)}$ as the corresponding change in $\eta_{\omega}$ when $\varphi_H = 0$ $(\frac{\pi}{2})$, and substituting $\Delta H_x = m_\mathrm{s} H_\mathrm{DL}$ and $\Delta H_y = - H_\mathrm{FL}$ into Eq.~(\ref{eq:torque:eta}), we have
\begin{equation}
\begin{aligned}
\label{eq:moke:sot}
&\begin{bmatrix}
\eta_{\omega,x}\\
m_\mathrm{s} \eta_{\omega,y}
\end{bmatrix}
= \frac{H}{H_k^2} \eta_{\mathrm{P}}\\
&\times \begin{bmatrix}
-\big(1 + r_\eta m_\mathrm{s} \sin (2 \varphi_E) \big) & -r_\eta \cos (2 \varphi_E)\\
r_\eta \cos (2 \varphi_E) & 1 - r_\eta m_\mathrm{s} \sin (2 \varphi_E)
\end{bmatrix}
\begin{bmatrix}
H_\mathrm{DL}\\
H_\mathrm{FL}
\end{bmatrix},\\
& \ \ \ \ \ \ \ \ \ r_\eta \equiv \frac{\eta_{\mathrm{Q}}}{\eta_{\mathrm{P}}}.
\end{aligned}
\end{equation}
Note that the determinant of the $2 \times 2$ matrix on the right hand side of Eqs.~(\ref{eq:hall:sot}) and (\ref{eq:moke:sot}) is $1 - r_H^2$ and $1 - r_\eta^2$, respectively. 
Equations~(\ref{eq:hall:sot}) and (\ref{eq:moke:sot}) thus indicate that $H_\mathrm{DL}$ and $H_\mathrm{FL}$ cannot be determined when $r_\mathrm{H}, r_\eta = \pm1$.
Such singularity was previously identified for the harmonic Hall voltage measurements: studies have shown that estimating $H_\mathrm{DL}$ and $H_\mathrm{FL}$ using the harmonic Hall resistances faces a problem when $r_\mathrm{H}$ is close to 1\cite{torrejon2014ncomm,hayashi2014prb}.
We infer that any signal in $R_{2\omega}$ that is not related to the current-induced torque, \textit{e.g.}, thermo-electric effects, is effectively amplified when $r_\mathrm{H}$ is close to 1.
Equations~(\ref{eq:hall:sot}) and (\ref{eq:moke:sot}) show that similar difficulty is expected for the optical measurements via MOKE.
In the following, we experimentally show that $r_\eta$ is significantly smaller than 1 even when $r_\mathrm{H}$ close to 1, allowing optical measurements to assess the current-induced torque accurately.

Before discussing the experimental results, we note an important feature of the optical measurements.
In contrast to the electrical measurements, the optical measurements offer an additional degree of freedom\cite{chen2017prb}, the polarization of the incident light.
Let us consider a case where we fix the polarization of the incident light along the $x$-axis.
Substituting $\varphi_E=0$, Eq.~(\ref{eq:moke:sot}) reduces to
\begin{equation}
\begin{aligned}
\label{eq:moke:sot:pol:x}
\begin{bmatrix}
\eta_{\omega,x}\\
m_\mathrm{s} \eta_{\omega,y}
\end{bmatrix}
&= \frac{H}{H_k^2} \eta_{\mathrm{P}}
\begin{bmatrix}
-1 &-r_\eta\\
r_\eta &1
\end{bmatrix}
\begin{bmatrix}
H_\mathrm{DL}\\
H_\mathrm{FL}
\end{bmatrix}.
\end{aligned}
\end{equation}
If the polarization is fixed along the $y$-axis ($\varphi_E = \frac{\pi}{2}$),
 Eq.~(\ref{eq:moke:sot}) reads
\begin{equation}
\begin{aligned}
\label{eq:moke:sot:pol:y}
\begin{bmatrix}
\eta_{\omega,x}\\
m_\mathrm{s} \eta_{\omega,y}
\end{bmatrix}
&= \frac{H}{H_k^2} \eta_{\mathrm{P}}
\begin{bmatrix}
-1 &r_\eta\\
-r_\eta &1
\end{bmatrix}
\begin{bmatrix}
H_\mathrm{DL}\\
H_\mathrm{FL}
\end{bmatrix}.
\end{aligned}
\end{equation}
As stated above, the determinant of the $2 \times 2$ matrix on the right hand side of Eqs.~(\ref{eq:moke:sot:pol:x}) and (\ref{eq:moke:sot:pol:y}) is zero when $r_\eta = \pm 1$.
We can change the polarization of the incident light when we change the magnetic field direction to measure $\eta_{\omega,x}$ or $\eta_{\omega,y}$.
For example, we can set the polarization along the $x$-axis when the magnetic field is parallel to the $y$-axis (to measure $\eta_{\omega,y}$), and then set the polarization along the $y$-axis when the field points along the $x$-axis (to determine $\eta_{\omega,x}$).
We therefore substitute $\varphi_E = 0$ for the lower row of the $2 \times 2$ matrix in Eq.~(\ref{eq:moke:sot}) and $\varphi_E = \frac{\pi}{2}$ for the upper row to obtain
\begin{equation}
\begin{aligned}
\label{eq:moke:sot:pol:xy}
\begin{bmatrix}
\eta_{\omega,x}\\
m_\mathrm{s} \eta_{\omega,y}
\end{bmatrix}
&= \frac{H}{H_k^2} \eta_{\mathrm{P}}
\begin{bmatrix}
-1 &r_\eta\\
r_\eta &1
\end{bmatrix}
\begin{bmatrix}
H_\mathrm{DL}\\
H_\mathrm{FL}
\end{bmatrix}.
\end{aligned}
\end{equation}
Now the determinant of the $2 \times 2$ matrix on the right-hand side is not zero when $r_\eta = \pm1$. 
It is thus possible to perform torque measurements even when $r_\eta$ is close to 1.
Note that the measurement scheme, whether one fixes the polarization direction or changes it with the magnetic field, does not influence the results if $r_\eta \ll 1$.

\section{Experimental results}

\subsection{\label{sec:samples}Samples}
NM/FM bilayers were prepared using radio frequency magnetron sputtering.
Thin films were deposited on oxidized silicon substrates at room temperature.
The thickness of silicon oxide was $\sim$100 nm, which increases the MOKE signal via optical interference effect\cite{sumi2018scirep,kimel2022jpd}.
The film structure is sub./2 Ta/1 CoFeB/2 MgO/1 Ta, referred to as Ta/CoFeB, and sub./3 W/1 CoFeB/2 MgO/1 Ta, referred to as W/CoFeB hereafter.
The unit of the film thickness is in nanometer.

Films were annealed at 300 $^\mathrm{o}$C to induce perpendicular magnetic anisotropy at the CoFeB/MgO interface\cite{ikeda2010nmat}.
Standard optical lithography and Ar ion milling were used to pattern the films into Hall bars.
The width $w$ and length $L$ of the current channel of the Hall bars are \SI{0.4}{mm} and \SI{1.2}{mm} (10 $\upmu$m and 25 $\upmu$m), respectively, for the optical (electrical) measurements. See Fig.~\ref{fig:schematic} for the definition of $w$ and $L$.

\subsection{Experimental setup}
The electrical transport properties of the Hall bars and the current-induced torque were measured using DC current and AC voltage sources, respectively.
For the former, a constant current $I_x$ is fed from the source, and the resulting Hall voltage $V_{yx}$ was measured using a voltmeter.
The Hall resistance is defined as $R_{yx} = V_{yx} / I_x $.
For the latter, the frequency of the AC voltage source is set to 2030 Hz, and the in-phase first harmonic $V_\omega$ and the (90 deg) out-of-phase second harmonic $V_{2\omega}$ Hall voltages were measured using a lock-in amplifier.
The current $I_x$ that passes through the Hall bar was measured using a series resistor and an AC voltmeter.
Definitions of the sign of $V_{xx}$ and $V_{yx}$ are illustrated in Fig.~\ref{fig:schematic}.

\begin{figure}[tb]
    \centering
    \includegraphics[width=1.0\linewidth]{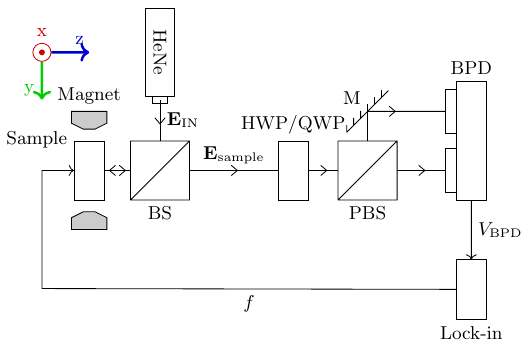}
    \caption{
Schematic drawing of the optical setup. BS: beam splitter, HWP: half-wave plate, QWP: quarter-wave plate, PBS: polarized beam splitter, BPD: balanced photodetector.
    }
    \label{fig:opticalsetup}
\end{figure}

For the optical measurements, the polarization angle ($ \theta_\mathrm{K} $) or the ellipticity ($ \eta_\mathrm{K} $) of the reflected light was measured.
Figure~\ref{fig:opticalsetup} shows the optical setup used in the experiments to extract $ \theta_\mathrm{K} $ and $ \eta_\mathrm{K}$.
The light source is a continuous-wave (CW) He-Ne laser.
The wavelength and the power of the laser were 633 nm and 5 mW, respectively.
The sample was irradiated with a linearly polarized light from a normal incidence.
In accordance to Eq.~(\ref{eq:moke:sot:pol:xy}), the polarization of the incident light is changed when the magnetic field direction is changed, that is, the polarization is parallel to the $x$-axis ($y$-axis) when the magnetic field is applied along the $y$-axis ($x$-axis).
The light reflected from the sample passes through a half-wave plate (HWP) or a quarter-wave plate (QWP).
The HWP or QWP is used to measure the polarization angle $\theta_\mathrm{K}$ or the ellipticity $ \eta_\mathrm{K} $ of the reflected light, respectively.
The light then enters a polarizing beamsplitter and a silicon-based balanced photodetector (BPD).
For the optical constant measurements (e.g., $\eta_\mathrm{P}$, $\eta_\mathrm{Q}$), the voltage from the BPD was detected using a voltmeter.
The field was swept while the voltmeter acquires signal.
We repeated the field sweep 300 - 500 times to improve the signal to noise ratio.
For the measurements of current-induced torque, an AC voltage source was supplied to the sample (Hall bars), and the in-phase first harmonic BPD voltage was measured using a lock-in amplifier.
The frequency of the AC voltage source was set to 2030 Hz.
The field or current were swept while the lock-in amplifier collected signal from the BPD. 
The sweep was repeated $\sim$25 times.
The current $I_x$ that passes through the sample is measured using a series resistor and an AC voltmeter.
The definition of the ellipticity $\eta_\mathrm{K}$, which equals the arctangent of the short and long axis of the ellipse formed by the light polarization, is drawn in Fig.~\ref{fig:schematic}.
Positive $\theta_\mathrm{K}$ and $\eta_\mathrm{K}$ corresponds to states with the polarization and elliptical angle of the reflected light rotate clockwise when seen from the sample side.

All measurements were performed at room temperature under ambient conditions.

\subsection{\label{sec:transport}Transport and optical constants}
\begin{figure}[tb]
\begin{center}
\includegraphics[width=1\linewidth]{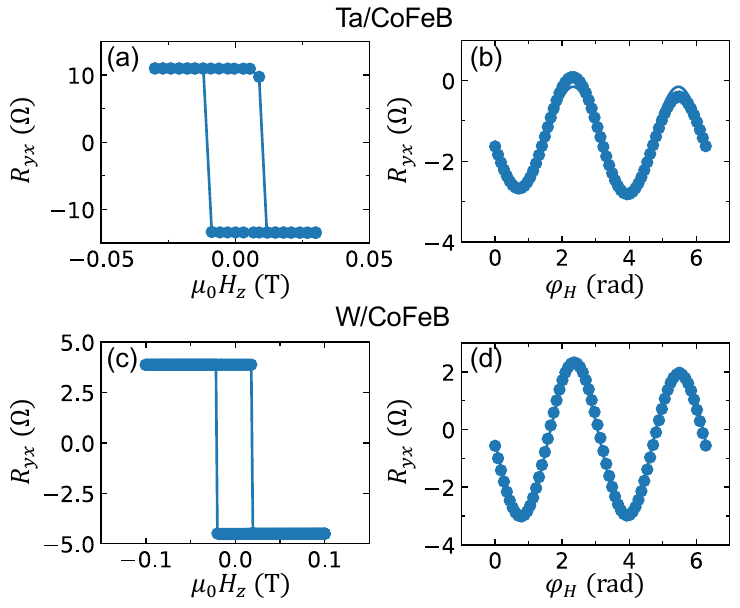}
    \caption{ (a,c) The out of plane field $H_z$ dependence of the Hall resistance $R_{yx}$. 
    (b,d) $R_{yx}$ plotted against the in-plane magnetic field angle $\varphi_H$. The field amplitude is 1.7 T. The solid lines show fit to the data with a sinusoidal function. (a,b) Ta/CoFeB, (c,d) W/CoFeB.
    }
    \label{fig:Ryx:ahephe}
\end{center}
\end{figure}

We swept the magnetic field along the film normal (along the $z$-axis) and measured the Hall resistance $R_{yx}$ to obtain the anomalous Hall resistance $R_\mathrm{A}$.
The out-of-plane field $H_z$ dependence of $R_{yx}$ is shown in Fig.~\ref{fig:Ryx:ahephe}(a,c) for Ta/CoFeB and W/CoFeB.
The difference of $R_{yx}$ between the two metastable states at zero magnetic field corresponds to $2 R_\mathrm{A}$.
The square hysteresis loop shows that the remanence is close to a single domain state. 

The planar Hall resistance $R_\mathrm{P}$ was obtained by rotating the magnetic field within the film plane.
$R_{yx}$ was measured against the in-plane field angle $\varphi_H$ while the field amplitude $H$ was fixed to 1.7 T.
Under these conditions ($H > H_k$), the magnetization rotates within the film plane, and its direction is aligned with that of the magnetic field.
The experimental results are presented in Fig.~\ref{fig:Ryx:ahephe}(b,d).
$R_{yx}$ varies with $\sin (2 \varphi_H)$, which is in agreement with Eq.~(\ref{eq:hall}).
The amplitude of the sinusoidal variation is equal to $R_\mathrm{P}$.
$R_\mathrm{P}$ is extracted from the results by fitting the data with a sinusoidal function.
Values of $R_\mathrm{A}$ and $R_\mathrm{P}$ are listed in Table~\ref{table:par}.

\begin{figure}[tb]
\begin{center}
   \includegraphics[width=\linewidth]{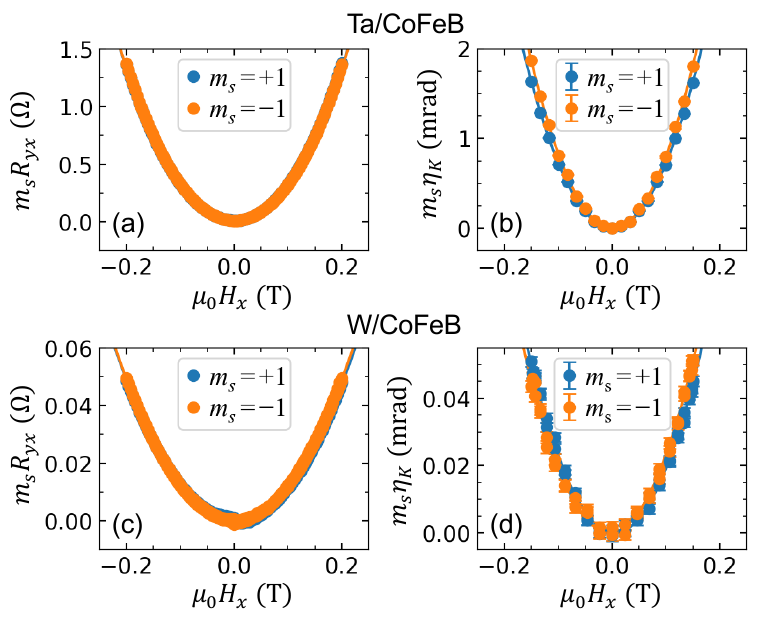} 
    \caption{ (a-d) The in-plane field $H_x$ dependence of the Hall resistance $R_{yx}$ (a,c) and Kerr ellipticity $\eta_\mathrm{K}$ (b,d). The solid lines represent fit to the data with a parabolic function. $R_{yx}$ here is obtained using the AC electrical measurements, that is, $R_{yx} = R_{\omega}$. The error bars in (b,d) represent the standard deviation of the MOKE signal over multiple measurements. (a,b) Ta/CoFeB, (c,d) W/CoFeB.
	}
    \label{fig:Ryx:Kerr:Hx}
    \end{center}
\end{figure}

Next, the Hall resistance $R_{yx}$ and the Kerr ellipticity $\eta_\mathrm{K}$ are measured as a function of the in-plane magnetic field to estimate $H_k$ and $\eta_\mathrm{P}$.
We sweep the amplitude of the in-plane field with a fixed field angle $\varphi_H = 0$.
The field is small enough such that its effect is to tilt the magnetization direction from the easy axis ($z$ axis).  
Figure~\ref{fig:Ryx:Kerr:Hx} shows the magnetic field ($H_x$) dependence of $R_{yx}$ and $\eta_\mathrm{K}$ for Ta/CoFeB and W/CoFeB. 
Symbol colors indicate the equilibrium magnetization direction: blue circles (orange circles) represent results when the equilibrium magnetization points along $+z$ ($-z$). 

The change in $R_{yx}$ and $\eta_\mathrm{K}$ against $H_x$ is parabolic, which is consistent with Eqs.~(\ref{eq:hall:thetadep}) and (\ref{eq:kerr:thetadep}).
The experimental data are fitted with a parabolic function.
The fitting results, presented by the solid lines in Fig.~\ref{fig:Ryx:Kerr:Hx}, describe the data well.
In Eqs.~(\ref{eq:hall:thetadep}) and (\ref{eq:kerr:thetadep}), we define the curvature $c_\mathrm{H}$ and $c_\mathrm{K}$ of the parabolic function as:
\begin{equation}
\label{eq:curvature:Hall}
    c_\mathrm{H} \equiv \dfrac{R_\mathrm{A} m_\mathrm{s} - R_\mathrm{P}\sin(2\varphi_H)}{2 H_k^2},
\end{equation}
\begin{equation}
\label{eq:curvature:eta}
    c_\mathrm{K} \equiv \dfrac{\eta_\mathrm{P} m_\mathrm{s} - \eta_\mathrm{Q}\sin(2\varphi_H)}{2 H_k^2}.
\end{equation}
Setting $\varphi_H = 0$ in Eqs.~(\ref{eq:curvature:Hall}) and (\ref{eq:curvature:eta}), $\frac{R_\mathrm{A}}{2 H_k^2}$ and $\frac{\eta_\mathrm{P}}{2 H_k^2}$ are extracted from the values of $c_\mathrm{H}$ and $c_\mathrm{K}$.
Using the value of $R_\mathrm{A}$ listed in Table~\ref{table:par}, we determine $H_k$ and $\eta_\mathrm{P}$.
The values are presented in Table~\ref{table:par}.

\begin{figure}[bt]
\begin{center}
\includegraphics[width=\linewidth]{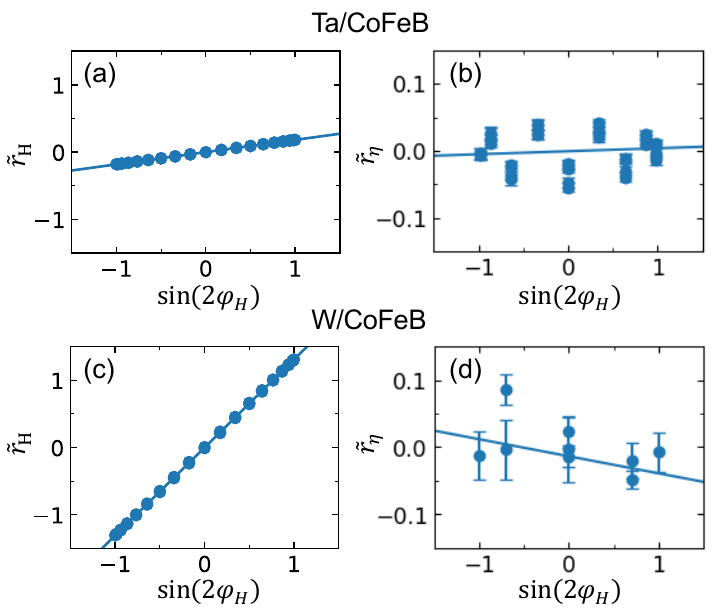}
    \caption{$\sin (2\varphi)$ dependence of $\tilde{r}_\mathrm{H}$ (a,c) and $\tilde{r}_\eta$ (b,d). (a,b) Ta/CoFeB, (c,d) W/CoFeB. The error bars in (b,d) represent errors of fitting Eq.~(\ref{eq:kerr:thetadep}) to the data.}
    \label{fig:r-paramter}
    \end{center}
\end{figure}

To determine $\eta_\mathrm{Q}$, the $H$ dependence of $\eta_\mathrm{K}$ is measured with various $\varphi_H$.
(For completeness, we also measure the $H$ dependence of $R_{yx}$ at different $\varphi_H$.)
The curvature of the parabolic function, $c_\mathrm{H}$ and $c_\mathrm{K}$, is obtained for each sweep at a fixed $\varphi_H$.
From Eq.~(\ref{eq:hall:thetadep}), (\ref{eq:kerr:thetadep}), (\ref{eq:curvature:Hall}) and (\ref{eq:curvature:eta}), we define
\begin{equation}
\begin{aligned}
\label{eq:hall:phe}
        \tilde{r}_\mathrm{H} \equiv - \left( \frac{2 H_k^2}{R_\mathrm{A}} c_\mathrm{H}  - m_s \right) = \frac{R_\mathrm{P}}{R_\mathrm{A}} \sin(2\varphi_H),
\end{aligned}
\end{equation}
\begin{equation}
\begin{aligned}
\label{eq:kerr:thetadep:a}
    \tilde{r}_\eta \equiv - \left( \frac{2 H_k^2}{\eta_\mathrm{P}} c_\mathrm{K}  - m_s \right) = \frac{\eta_\mathrm{Q}}{\eta_\mathrm{P}} \sin(2\varphi_H).
\end{aligned}
\end{equation}
At a given angle $\varphi_H$, we multiply $\frac{2 H_k^2}{R_\mathrm{A}}$ ($\frac{2 H_k^2}{\eta_\mathrm{P}}$) to $c_\mathrm{H}$ ($c_\mathrm{K}$) obtained from the experiments and subtract $m_s$ to obtain $\tilde{r}_\mathrm{H}$ and $\tilde{r}_\eta$.
In Fig.~\ref{fig:r-paramter}, we plot the $\sin(2\varphi_H)$ dependence of $\tilde{r}_\mathrm{H}$ and $\tilde{r}_\eta$.
As is evident, for both films (Ta/CoFeB and W/CoFeB), $\tilde{r}_\mathrm{H}$ linearly scales with $\sin(2\varphi_H)$ whereas $\tilde{r}_\eta$ is almost constant.
We fit the data with a linear function.
The slope of the linear function is equal to $\frac{R_\mathrm{P}}{R_\mathrm{A}}$ for $\tilde{r}_\mathrm{H}$ vs. $\sin(2\varphi_H)$ [Fig.~\ref{fig:r-paramter}(a,c)] and $\frac{\eta_\mathrm{Q}}{\eta_\mathrm{P}}$ for $\tilde{r}_\eta$ vs. $\sin(2\varphi_H)$ [Fig.~\ref{fig:r-paramter}(b,d)].
From the slope and $\eta_\mathrm{P}$ listed in Table~\ref{table:par}, we estimate $\eta_\mathrm{Q}$, which is presented in Table~\ref{table:par}.
$R_\mathrm{P}$ obtained from the slope and $R_\mathrm{A}$ is in good agreement with that determined from the results shown in Fig.~\ref{fig:Ryx:ahephe}(b,d).

Table~\ref{table:par} summarizes the parameters obtained in the experiments.
In both films, $|\eta_\mathrm{P}|$ is significantly larger than $|\eta_\mathrm{Q}|$.
In contrast, $|R_\mathrm{A}|$ is considerably larger than $|R_\mathrm{P}|$ for Ta/CoFeB whereas $|R_\mathrm{P}|$ is larger than $|R_\mathrm{A}|$ for W/CoFeB.
The ratio of the Hall resistances $r_\mathrm{H} = \frac{R_\mathrm{P}}{R_\mathrm{A}}$ and the MOKE components $r_\eta = \frac{\eta_\mathrm{Q}}{\eta_\mathrm{P}}$, defined in Eqs.~(\ref{eq:hall:sot}) and (\ref{eq:moke:sot}), respectively, are also presented in Table~\ref{table:par}.
Whereas $r_\eta$ is considerably smaller than 1 for both structures, $r_\mathrm{H} < 1 $ for Ta/CoFeB but $r_\mathrm{H} \sim 1$ for W/CoFeB.
The large $r_\mathrm{H}$ in W/CoFeB causes a significant difficulty in extracting spin-torque efficiency, as discussed in the next section.
\begin{table}[tb]
    \caption{Summary of the parameters obtained from the experiments. Effective magnetic anisotropy field $H_k$, anomalous Hall resistance $R_\mathrm{A}$, planar Hall resistance $R_\mathrm{P}$, Hall resistance ratio $r_\mathrm{H} = R_\mathrm{P} / R_\mathrm{A}$, polar $\eta_\mathrm{P}$ and quadratic $\eta_\mathrm{Q}$ components of the Kerr ellipticity, ratio of the quadratic to polar components of Kerr ellipticity $r_\eta = \eta_\mathrm{Q} / \eta_\mathrm{P}$ for Ta/CoFeB and W/CoFeB bilayers are presented. For a reference, data from Refs.~\cite{kim2016prl} and \cite{fan2016apl} are included. The error range for $r_\eta$ represents the errors associated with fitting a linear function to the data $r_\eta$ vs. $\sin(2 \varphi_H)$ shown in Fig.~\ref{fig:r-paramter}(b,d).
    }
    \label{table:par}
    \centering
    \begin{tabular}{c c c c c c c c}
    \toprule
       parameter  & $H_k$ & $R_\mathrm{A}$ & $R_\mathrm{P}$ & $r_\mathrm{H}$ & $\eta_\mathrm{P}$ & $\eta_\mathrm{Q}$ & $r_\eta$\\
     unit & $T$ & $\Omega$ & $\Omega$ &  & mrad & mrad &  \\
    \midrule
   Ta/CoFeB & 0.42 & -12 & -2.6 & 0.22 & -27 & -0.08 & $0.003 \pm 0.01$\\
   W/CoFeB & 1.3 & -4.2 & -5.1 & 1.2 & -7.2 & 0.2 & -$0.03 \pm 0.02$\\
   CoFeB\footnote{Ref.~\cite{kim2016prl}: 0.1 Ta/1 CoFeB/2 MgO/1 Ta, 300 $^{o}$C annealed} &  & -39 & -4 & 0.10 & &  & \\
   Pt/Py\footnote{Ref.~\cite{fan2016apl}: 6 Pt/8 NiFe/AlO$_{x}$} &  &  &  & & 5.8 & 0.01 & 0.002\\
    \bottomrule
    \end{tabular}
\end{table}

\subsection{Current-induced torque}
In this section, we discuss the results of the current-induced torque.
An AC voltage (amplitude: $V_0$, frequency: 2030 Hz) was supplied to the bilayer, and the resulting second harmonic Hall voltage $V_{2 \omega}$ or the first harmonic MOKE signal $ \eta_{\omega} $ was measured.
The current density $j$ that flows into the NM layer is calculated assuming a parallel resistor circuit model.
The resistivity of each layer is obtained from that reported in similar systems (Ta: \SI{200}{\micro\ohm\cm}, W: \SI{110}{\micro\ohm\cm}, CoFeB: \SI{160}{\micro\ohm\cm}).
For optical measurements, we limited the maximum $j$ to \SI{1e9}{A/m^{2}} to avoid parasitic current-induced heating effects\cite{marui2018apex,riego2016apl,su2017apl,stamm2017prl}.
An in-plane magnetic field $\bm{H}$ is applied during the measurements.

\begin{figure}[b]
    \centering
    \includegraphics[width=\linewidth]{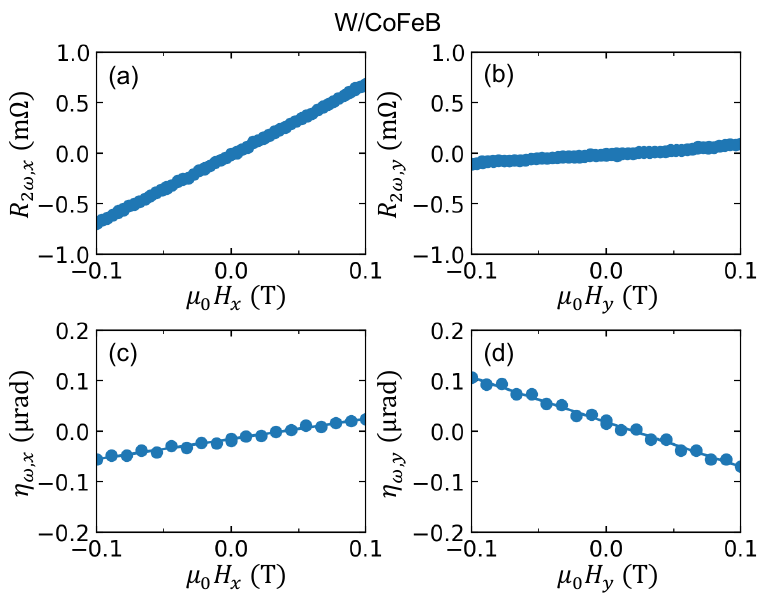}
    \caption{
Magnetic field dependence of the second harmonic Hall resistance $R_{2 \omega}$ (a,b) and the Kerr ellipticity $\eta_{\omega}$ (c,d) for W/CoFeB. The field angle $\varphi_H $ is set to 0 deg (a,c) and 90 deg (b,d). The AC voltage is set to 10 V$_\mathrm{rms}$ for the second harmonic Hall resistance measurements and 5 V$_\mathrm{rms}$ for the MOKE measurements. The equilibrium magnetization direction points along $+z$.
    }
    \label{fig:torque}
\end{figure}

The dependence on the in-plane field ($H$) of $R_{2 \omega}$ for W/CoFeB is plotted in Fig.~\ref{fig:torque}(a,b) when $\varphi_H = 0$ deg (denoted as $R_{2 \omega,x}$) and $\varphi_H = 90$ deg ($R_{2 \omega,y}$).
Here, the equilibrium magnetization direction is set along $+z$, \textit{i.e.} $m_s = 1$.
Both $R_{2 \omega, x}$ and $R_{2 \omega, y}$ vary almost linearly with $H_{x(y)}$.
We fit the data with a linear function to obtain the slope of $R_{2 \omega, x(y)}$ vs. $H$, which is substituted into Eq.~(\ref{eq:hall:sot}) to estimate the current-induced torque $H_\mathrm{DL}$ and $H_\mathrm{FL}$.

Figures~\ref{fig:torque}(c) and \ref{fig:torque}(d) show the $H$ dependence of $\eta_{\omega}$
for W/CoFeB when the field angle $\varphi_H$ is set to 0 deg (denoted as $\eta_{\omega,x}$) and 90 deg ($\eta_{\omega,y}$), respectively.
$\eta_{\omega,x}$ and $\eta_{\omega,y}$ also vary linearly with $H$ in this range.
We therefore fix $H$ to $\pm 0.1$ T for the optical measurements and calculate $\eta_{\omega, x(y)}/H$ to estimate $H_\mathrm{DL}$ and $H_\mathrm{FL}$ using Eq.~(\ref{eq:moke:sot}).
See Appendix Sec.~\ref{sec:jdep} for the $j$ dependence of $\eta_{\omega, x(y)}$.

In estimating the effective field $H_\mathrm{DL(FL)}$ from $R_{2\omega,x(y)}$ and $\eta_{\omega,x(y)}$ using Eq.~(\ref{eq:hall:sot}) and Eq.~(\ref{eq:moke:sot}), values of $r_\mathrm{H}$, $R_\mathrm{A}$, $r_\eta$, $\eta_\mathrm{P}$ and $H_k$ listed in Table~\ref{table:par} are used.
From $H_\mathrm{DL(FL)}$, the damping-like (field-like) spin-torque efficiency $\xi_\mathrm{DL(FL)}$ is calculated using the following relations:
\begin{equation}
\begin{aligned}
\label{eq:spin-torque-efficiency}
    \xi_\mathrm{DL} &= \dfrac{(H_\mathrm{DL}/j) \times M_\mathrm{s} t} {\hbar/2e},\\
    \xi_\mathrm{FL} &= \dfrac{(H_\mathrm{FL}/j) \times M_\mathrm{s} t} {\hbar/2e},
\end{aligned}
\end{equation}
where $M_\mathrm{s}$ is the saturation magnetization and $t$ is the thickness of the FM layer.
$\hbar$ is the reduced Planck constant and $e$ is the elementary charge.
Here, we assume a transparent NM/FM interface for spin transport: $\xi_\mathrm{DL}$ and $\xi_\mathrm{FL}$ therefore provide the lower limit of the spin-torque efficiency.
The values of $\xi_\mathrm{DL}$ and $\xi_\mathrm{FL}$, obtained from the electrical and optical measurements for Ta/CoFeB and W/CoFeB, are summarized in Table~\ref{table:sot}.
We assume $M_\textrm{s} = 900$ kA/m for Ta/CoFeB and $M_\textrm{s} = 1000$ kA/m for W/CoFeB\cite{lau2017jjap}.
\begin{table}[b]
    \caption{Damping-like $\xi_\mathrm{DL}$ and field-like $\xi_\mathrm{FL}$ spin-torque efficiencies obtained from the harmonic Hall voltage measurements and the MOKE measurements for Ta/CoFeB and W/CoFeB.
    }
    \label{table:sot}
    \centering
    \begin{tabular}{c c c c c}
    \toprule
     & \multicolumn{2}{c}{Hall} & \multicolumn{2}{c}{MOKE}\\
     & $\xi_\mathrm{DL}$ & $\xi_\mathrm{FL}$ & $\xi_\mathrm{DL}$ & $\xi_\mathrm{FL}$\\
     \midrule 
   Ta/CoFeB & -0.16 & 0.30 & -0.10 & 0.16\\
   W/CoFeB & 4.0 & -4.2 & -0.18 & 0.36\\
     \bottomrule
     \end{tabular}
\end{table}

For Ta/CoFeB, $\xi_\mathrm{DL}$ and $\xi_\mathrm{FL}$ obtained from the electrical and optical measurements show similar values.
In contrast, for W/CoFeB, the magnitude and even the sign of the spin-torque efficiency are different between the two approaches (electrical vs. optical).
As demonstrated in previous studies, the spin-torque efficiency estimated using electrical measurements returns an unphysical value\cite{torrejon2014ncomm,karimeddiny2023sciadv} when $r_\mathrm{H}$ is close to 1.
This behavior is clearly observed for the electrical measurements in W/CoFeB.
It is evident from Eq.~(\ref{eq:hall:sot}) [Eq.~(\ref{eq:moke:sot})] that one cannot extract the effective field $H_\mathrm{DL}$ and $H_\mathrm{FL}$ from the measured $R_{2\omega}$ [$\eta_{\omega}$] when $r_\mathrm{H}$ [$r_\eta$] is equal to 1, \textit{i.e.} when the determinant of the matrix in Eq.~(\ref{eq:hall:sot}) [Eq.~(\ref{eq:moke:sot})] is zero.
The conversion process thus seems to fail when $r_\mathrm{H}$ or $r_\eta$ approaches 1.
We infer that any signal in $R_{2\omega}$ and $\eta_{\omega}$ that is not related to the current-induced torque, \textit{e.g.}, thermo-electric effects, is effectively amplified when $r_\eta$ ($r_\mathrm{H}$) is close to 1, causing the inaccurate estimation.
For the optical measurements, interestingly, $r_\eta$ is significantly smaller than 1 even for W/CoFeB.
The estimated spin-torque efficiency, therefore, agrees with that reported in the literature\cite{sinova2015rmp,hoffmann2013ieee} for both Ta/CoFeB and W/CoFeB. 
Note that the choice of the geometry of the optical current-induced torque measurements, whether the incident light polarization is changed when we change the magnetic field direction [see the discussion along Eqs.~(\ref{eq:moke:sot:pol:x} - \ref{eq:moke:sot:pol:xy})] does not factor in here since $r_\eta \ll 1$. 

\section{Discussion}
The results described in the previous section shows that the relative size of the planar Hall to anomalous Hall resistances ($r_\mathrm{H}$) and the quadratic to polar components of the magneto-optical Kerr effect ($r_\eta$) are essential in determining the current-induced torque. 
For Ta/CoFeB, both $r_\mathrm{H}$ and $r_\eta$ are small, whereas $r_\mathrm{H}$ is close to 1 but $r_\eta$ is small in W/CoFeB.
It has been reported that the large $r_\mathrm{H}$ in W/CoFeB is due to the SMR\cite{kim2016prl}.
That is, the planar Hall resistance includes contributions from the anisotropic magnetoresistance and the SMR.
The SMR also contributes to the anomalous Hall resistance: $R_\mathrm{A}$ scales with the product of the SMR and the imaginary component of the spin mixing conductance\cite{nakayama2013prl,chen2013prb}; the latter can vary in magnitude and sign depending on the interface\cite{stiles2002prb,weiler2013prl,kim2014prb,sheng2017sciadv,zhu2019prl}.
As the SMR scales with the square of the spin Hall angle\cite{nakayama2013prl,chen2013prb}, materials with large spin Hall effect tend to exhibit large $r_\mathrm{H}$.
In Table~\ref{table:par}, we included $r_\mathrm{H}$ of a system that can be considered a single CoFeB layer with similar thickness used in this study\cite{kim2016prl}.
The value ($r_\mathrm{H} = 0.1$) is nearly half of that of Ta/CoFeB, highlighting contributions from the SMR is minor for Ta/CoFeB but large for W/CoFeB.

Most importantly, we find that $r_\eta$ is significantly smaller than 1 for both Ta/CoFeB and W/CoFeB.
That is, in contrast to the planar Hall resistance, the quadratic component of MOKE does not seem to be influenced by the SMR.
It should be noted that first-principles calculations indicate that the AC spin Hall conductivity at a light energy of 1.96 eV (corresponding to a light wavelength of 633 nm) is similar in magnitude with its DC counterpart: the former is roughly a factor of 2 smaller than the latter (data not shown here): see \textit{e.g.} Ref.~\cite{marui2023prb} for the DC and AC spin Hall conductivities of Pt.
Thus, the magnitudes of spin current generated from the DC and light electric field via the spin Hall effect are close.
We attribute the smallness of $r_\eta$ to the slow response of the diffusive spin current.
That is, under a DC driving force (electric field), sufficient time is available for the spin current to travel to the NM/FM interface and being reflected to induce magnetization-dependent charge current via the spin Hall and inverse spin Hall effects.
In contrast, time is not enough for the electrons to travel to the interface under the light electric field driving, limiting contribution from the SMR on the quadratic Kerr effect.

As a reference, we show in Table~\ref{table:par} the value of $r_\eta$ for Pt/NiFe, reported in Ref.~\cite{fan2016apl}.
$r_\eta$ for Pt/NiFe is also significantly smaller than 1. 
As the thickness of the NiFe layer is sufficiently large in Ref.~\cite{fan2016apl}, the influence of Pt on the polar and quadratic MOKE is small, and the value of $r_\eta$ reflects the value of NiFe: contribution from the SMR is negligible. 
These results thus suggest that $r_\eta$ of Ta/CoFeB and W/CoFeB is close to that of bulk ferromagnet.
In the Appendix, Sec.~\ref{sec:interference}, we discuss the influence of optical interference within the sample that may affect estimation of $r_\eta$.
We find that $r_\eta$ is significantly smaller than 1 in W/CoFeB even when the interference effects are taken into account.

\section{Conclusion}
We have studied the off-diagonal components of the conductivity tensor, both in the zero frequency limit and at a light frequency (474 THz) corresponding to a light wavelength of 633 nm, in Ta/CoFeB and W/CoFeB bilayers. 
We determine the planar Hall and anomalous Hall resistances using transport measurements and the quadratic and polar components of the magneto-optical Kerr effect (MOKE) via optical measurements.
The quadratic to polar MOKE ratio in both systems is significantly smaller than 1. In contrast, the planar Hall to anomalous Hall resistance ratio is close to 1 in W/CoFeB and equal to 0.2 in Ta/CoFeB.
The large planar Hall to anomalous Hall resistances ratio in W/CoFeB causes difficulty in estimating the spin-torque efficiency using the harmonic Hall voltage measurement.
In contrast, the small quadratic to polar MOKE ratio in both systems allows accurate determination of the spin-torque efficiency using MOKE.
The large planar Hall to anomalous Hall resistances ratio in W/CoFeB is known to be caused by spin transport in the W layer, \textit{i.e.} it is influenced by the spin Hall magnetoresistance.
The small quadratic to polar MOKE ratio in W/CoFeB thus suggests that the influence of spin current on the AC conductivity tensor is negligible, likely due to the fast driving force (i.e. light electric field) with respect to the time required for spin transport.
These findings demonstrate why MOKE serves as a versatile approach to determining spin-torque efficiency in non-magnetic metal/ferromagnetic metal bilayers.

\section{Appendix}
\subsection{\label{sec:jdep} Current density dependence of $\eta_{\omega, x(y)}$}
\begin{figure}[tb]
    \centering
    \includegraphics[width=\linewidth]{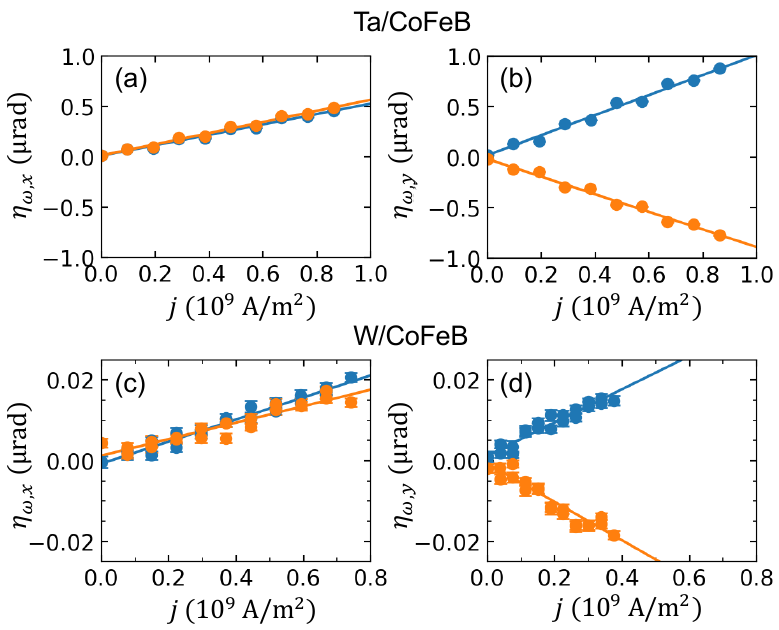}    
    \caption{
The current density $j$ dependence of the MOKE signal with magnetic field $H$ fixed to $0.1$ T for Ta/CoFeB (a,b) and W/CoFeB (c,d). The field angle $\varphi_H $ is set to 0 deg (a,c) and 90 deg (b,d). 
Blue and orange symbols represent data with $m_\mathrm{s} = 1$ and -1, respectively. Solid lines show a linear fit to the data.
    }
    \label{fig:torque:kerr:j}
\end{figure}

The current density $j$ dependence of $\eta_{\omega, x(y)}$, obtained at $H = 0.1$ T, is shown in Fig.~\ref{fig:torque:kerr:j} for Ta/CoFeB and W/CoFeB.
Different symbols indicate $m_s = \pm1$, \textit{i.e.}, the equilibrium magnetization direction points along $\pm z$.
In all cases, $\eta_{\omega, x(y)}$ varies linearly with $j$.
$\eta_{\omega, y}$ changes its sign when the equilibrium magnetization direction is reversed,
while the sign is the same for $\eta_{\omega, x}$.
These results are consistent with the $m_s$ dependence of $\eta_{\omega, x(y)}$: see Eqs.~(\ref{eq:moke:sot:pol:x} - \ref{eq:moke:sot:pol:xy}).

\subsection{\label{sec:interference} Optical interference effect}
\subsubsection{Model calculations}
In the main text, we showed the ratio $r_\eta = \frac{\eta_\mathrm{Q}}{\eta_\mathrm{P}}$ is significantly smaller than its electrical counterpart $r_\mathrm{H} = \frac{R_\mathrm{P}}{R_\mathrm{A}}$, allowing accurate optical measurements of the spin-torque efficiency.
We assumed that the smallness of $r_\eta$ in the same structure results from the absence of such contribution.
However, the magnitude of $r_\eta$ (and $r_\theta$) can be affected by the film stacking, including the substrate used.
In particular, if a dielectric layer is placed within the sample, optical interference effect can significantly enhance or suppress the MOKE signal.
This is indeed the case for the film used in the experiments: a SiO$_2$ layer is sandwiched between the Si substrate and the metallic films. 
In this section, we present numerical calculations of a model system to clarify the effect of the optical interference on $r_\eta = \frac{\eta_\mathrm{Q}}{\eta_\mathrm{P}}$ and $r_\theta = \frac{\theta_\mathrm{Q}}{\theta_\mathrm{P}}$.
We then discuss how to exclude the interference effect to extract the \textit{intrinsic} $r_\theta$ and $r_\eta$ of the film stack under study.

The structure of the model system is similar to that of the experiments: Si sub./$t_{\mathrm{SiO}_2}$ SiO$_2$/$t_\mathrm{FM}$ ferromagnetic metal (FM)/air.
The thickness of the substrate, SiO$_2$ and FM are set to 50 $\upmu$m, $t_{\mathrm{SiO}_2}$ and $t_\mathrm{FM}$, respectively.
The film, facing air, is irradiated with a linearly polarized light from the film normal.
We calculate the MOKE signal using the approach described in Refs.~\cite{yeh1980surfsci,visnovsky1986czechjpba}.
Specifically, the Maxwell's equations are solved under the appropriate boundary conditions.
The following permittivity tensor is assumed for all media\cite{visnovsky1986czechjpbb,buchmeier2009prb}. 
\begin{widetext}
\begin{equation}
    \dfrac{\epsilon}{\epsilon_{0}} = n^2
    \begin{bmatrix}
         1 & 0 & 0\\
        0 & 1 & 0\\
        0 & 0 & 1\\
     \end{bmatrix}
    + P 
    \begin{bmatrix}
        0 & m_z & -m_y\\
        -m_z & 0 & m_x\\
        m_y & -m_x & 0\\        
    \end{bmatrix}
    + 
    \begin{bmatrix}
        (G_{12} - G_{11}) ( m_y^2 + m_z^2) & G_{44} m_x m_y & G_{44} m_x m_z\\
        G_{44} m_y m_x & (G_{12} - G_{11}) ( m_z^2 + m_x^2) & G_{44} m_y m_z\\
        G_{44} m_z m_x & G_{44} m_z m_y & (G_{12} - G_{11}) ( m_z^2 + m_x^2)\\        
    \end{bmatrix}
    \label{eq:epsilon}
\end{equation}
\end{widetext}
The refractive index $n$ of each medium are summarized in Table~\ref{table:n}. 
For the substrate, SiO$_2$ and air, we set $P=0$, $G_{11} = G_{12} = G_{44} = 0$.
The magneto-optical constants ($P, G_{ij}$) for the FM layer are presented in Table~\ref{tab:MOKE:FM}.
Here we used values of Co$_{20}$Fe$_{60}$B$_{20}$ for $P$\cite{sumi2018scirep} and Fe for $G_{ij}$\cite{buchmeier2009prb}.
To calculate $\theta_\mathrm{P}$ and $\eta_\mathrm{P}$, we turn off contribution from the quadratic component of the dielectric tensor, \textit{i.e.} $G_{ij} = 0$, and set $(m_x, m_y, m_z) = (0,0,1)$, whereas $\theta_\mathrm{Q}$ and $\eta_\mathrm{Q}$ are obtained by setting the polar component $P$ to zero and $(m_x, m_y, m_z) = \frac{1}{\sqrt{2}} (1,1,0)$.
The film normal is set parallel to the $z$-axis.

\begin{table}[b]
    \caption{Refractive index $n$ of the materials used in the model system.}
    \centering
    \begin{tabular}{ccc}
        \toprule 
         material & value & source\\
        \midrule
        Air & $1.0$ & \\
        FM & $2.448 + 2.758 i$ & Ref.~\cite{sumi2018scirep}\\
        SiO$_2$ & $1.5$ & \\
        Si & $3.706 + 0.123 i$ & Ref.~\cite{sumi2018scirep}\\
        \bottomrule
    \end{tabular}
    \label{table:n}
\end{table}

\begin{table}[tb]
    \caption{MOKE parameters of the FM layer.}
    \centering
    \begin{tabular}{ccc}
        \toprule 
         parameter & value & source\\
        \midrule
        $P$ & $0.149 - 0.274 i$ & Ref.~\cite{sumi2018scirep}\\
        $G_{11} - G_{12}$ & $- 0.0358 - 0.0382 i$ & Ref.~\cite{buchmeier2009prb}\\
        $G_{44}$ & $- 0.0117 - 0.0349 i$ & Ref.~\cite{buchmeier2009prb}\\
        \bottomrule
    \end{tabular}
    \label{tab:MOKE:FM}
\end{table}

\begin{figure}[b]
    \centering
    \includegraphics[width=1.0\linewidth]{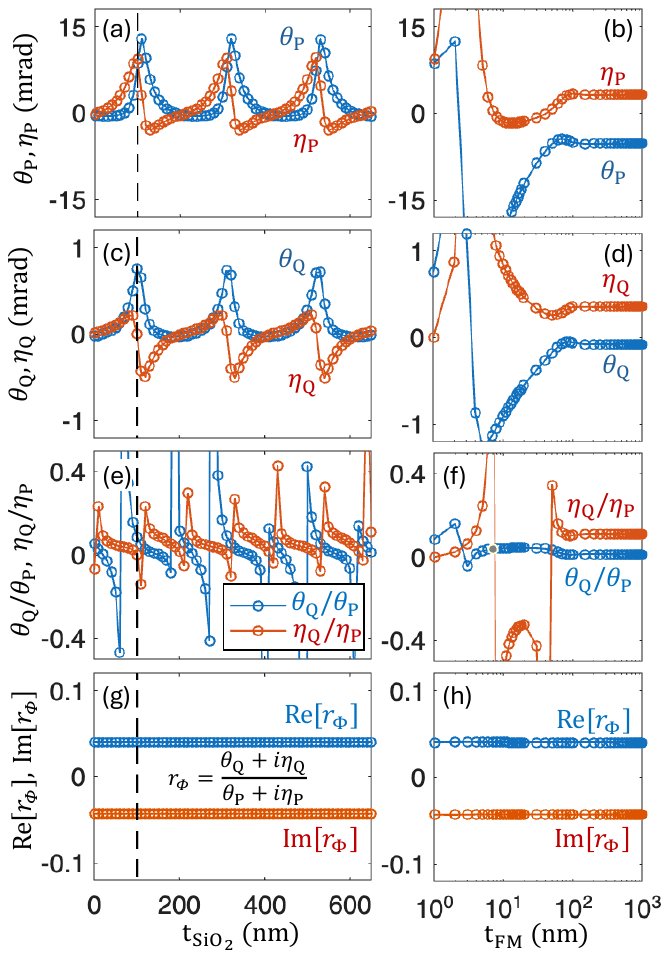}
    \caption{(a-d) Calculated MOKE signals $\theta_\mathrm{P}$, $\eta_\mathrm{P}$ (a,b) and $\theta_\mathrm{Q}$, $\eta_\mathrm{Q}$ (c,d) plotted as a function of SiO$_{2}$ thickness $t_{\mathrm{SiO}_2}$ (a,c) and FM layer thickness $t_\mathrm{FM}$ (b,d). (e,f) $t_{\mathrm{SiO}_2}$ (e) and $t_\mathrm{FM}$ (f) dependence of $\frac{\theta_\mathrm{Q}}{\theta_\mathrm{P}}$, $\frac{\eta_\mathrm{Q}}{\eta_\mathrm{P}}$. (g,h) Real and imaginary parts of $r_\Phi = \frac{\theta_\mathrm{Q} + i \eta_\mathrm{Q}}{\theta_\mathrm{P} + i \eta_\mathrm{P}}$  vs. $t_{\mathrm{SiO}_2}$ (g) and $t_\mathrm{FM}$ (h). The vertical dashed line in (a,c,e,g) indicates $t_\mathrm{SiO2}$ used in the experiments.}
    \label{fig:sio2dep}
\end{figure}

The calculated polar ($\theta_\mathrm{P}$, $\eta_\mathrm{P}$) and quadratic ($\theta_\mathrm{Q}$, $\eta_\mathrm{Q}$) MOKE signals are plotted against $t_{\mathrm{SiO}_2}$ and $t_\mathrm{FM}$ in Figs.~\ref{fig:sio2dep}(a,c) and \ref{fig:sio2dep}(b,d), respectively.
The blue and red circles represent the real ($\theta_\mathrm{P}$, $\theta_\mathrm{Q}$) and imaginary ($\eta_\mathrm{P}$, $\eta_\mathrm{Q}$) parts of the MOKE signal.
As is evident from the plots shown in Fig.~\ref{fig:sio2dep}(a,c), all signals show strong dependence on $t_{\mathrm{SiO}_2}$.
The extrema found in the plots are caused by multiple reflections within the SiO$_2$ layer\cite{sumi2018scirep,kimel2022jpd}.
The $t_\mathrm{FM}$ dependence of the MOKE signals [Fig.~\ref{fig:sio2dep}(b,d)] shows a large variation when $t_\mathrm{FM}$ is small, but tends to converge as the thickness is increased (note the horizontal axis is in log-scale).
The MOKE signal in the large $t_\mathrm{FM}$ limit represents its \textit{intrinsic} value of the FM layer, which we denote using a superscript "s", \textit{i.e.} $\theta_\mathrm{Q}^\mathrm{s}$, $\theta_\mathrm{P}^\mathrm{s}$, $\eta_\mathrm{Q}^\mathrm{s}$, $\eta_\mathrm{P}^\mathrm{s}$.

The $t_{\mathrm{SiO}_2}$ and $t_\mathrm{FM}$ dependence of $r_\theta = \frac{\theta_\mathrm{Q}}{\theta_\mathrm{P}}$ and $r_\eta = \frac{\eta_\mathrm{Q}}{\eta_\mathrm{P}}$ are shown in Fig.~\ref{fig:sio2dep}(e,f).
$r_\theta$ and $r_\eta$ diverge at certain layer thicknesses when $\theta_\mathrm{P}$ or $\eta_\mathrm{P}$ crosses zero.
The optical interference therefore causes fluctuation in $r_\theta$ and $r_\eta$ around the intrinsic value of the FM layer, $r_\theta^\mathrm{s} = \frac{\theta_\mathrm{Q}^\mathrm{s}}{\theta_\mathrm{P}^\mathrm{s}}$ and $r_\eta^\mathrm{s} = \frac{\eta_\mathrm{Q}^\mathrm{s}}{\eta_\mathrm{P}^\mathrm{s}}$.
We note that, for spin-torque efficiency measurements, it is not necessary to know the values of $r_\theta^\mathrm{s}$ or $r_\eta^\mathrm{s}$ as long as $r_\theta$ or $r_\eta$ is known.
Here, however, we are interested in the value of $r_\theta^\mathrm{s}$ and $r_\eta^\mathrm{s}$ of the film stack to study whether there is a physical difference with $r_\mathrm{H}$.

To extract $r_\theta^\mathrm{s}$ and $r_\eta^\mathrm{s}$ of a given film stack, 
we use the following properties of the MOKE signal.
First, we define a quantity 
\begin{equation}
\begin{aligned}
\label{eq:rphi}
r_\Phi = \frac{\theta_\mathrm{Q} + i \eta_\mathrm{Q}}{\theta_\mathrm{P} + i \eta_\mathrm{P}}.
\end{aligned}
\end{equation}
In Fig.~\ref{fig:sio2dep}(g,h), we plot the thickness dependences of the real and imaginary parts of $r_\Phi$.
As is evident, these quantities are constant against $t_{\mathrm{SiO}_2}$ and $t_\mathrm{FM}$: they are not influenced by the optical interference effects.
We define $r_\theta^\mathrm{m} = \frac{\theta_\mathrm{Q}^\mathrm{m}}{\theta_\mathrm{P}^\mathrm{m}}$ and $r_\eta^\mathrm{m} = \frac{\eta_\mathrm{Q}^\mathrm{m}}{\eta_\mathrm{P}^\mathrm{m}}$ as the value of $r_\theta$ and $r_\eta$ for a given film stack obtained in the experiments.
In addition, we set $r_\Phi^\mathrm{s} \equiv \frac{\theta_\mathrm{Q}^\mathrm{s} + i \eta_\mathrm{Q}^\mathrm{s}}{\theta_\mathrm{P}^\mathrm{s} + i \eta_\mathrm{P}^\mathrm{s}}$ and $r_\Phi^\mathrm{m} \equiv \frac{\theta_\mathrm{Q}^\mathrm{m} + i \eta_\mathrm{Q}^\mathrm{m}}{\theta_\mathrm{P}^\mathrm{m} + i \eta_\mathrm{P}^\mathrm{m}}$.
From the results shown in Fig.~\ref{fig:sio2dep}(g,h), we have $\mathrm{Re}[r_\Phi^\mathrm{m}] = \mathrm{Re}[r_\Phi^\mathrm{s}]$ and $\mathrm{Im}[r_\Phi^\mathrm{m}] = \mathrm{Im}[r_\Phi^\mathrm{s}]$.
Using these relations, we obtain the following form:
\begin{equation}
\begin{aligned}
\label{eq:rs}
r_\theta^\mathrm{s} = \mathrm{Re}[r_\Phi^\mathrm{m}] - \frac{\eta_\mathrm{P}^\mathrm{s}}{\theta_\mathrm{P}^\mathrm{s}} \mathrm{Im}[r_\Phi^\mathrm{m}], \\
r_\eta^\mathrm{s} = \mathrm{Re}[r_\Phi^\mathrm{m}] + \frac{\theta_\mathrm{P}^\mathrm{s}}{\eta_\mathrm{P}^\mathrm{s}} \mathrm{Im}[r_\Phi^\mathrm{m}], \\
\end{aligned}
\end{equation}
Equation~(\ref{eq:rs}) suggests that one can estimate $r_\theta^\mathrm{s}$ and $r_\eta^\mathrm{s}$ from the measured value of $r_\Phi^\mathrm{m}$ for a given film stack if the ratio $\frac{\eta_\mathrm{P}^\mathrm{s}}{\theta_\mathrm{P}^\mathrm{s}}$ can be guessed.
In the next section, we show measurement results of real ($\theta_\mathrm{P}^\mathrm{m}$, $\theta_\mathrm{Q}^\mathrm{m}$) and imaginary ($\eta_\mathrm{P}^\mathrm{m}$, $\eta_\mathrm{Q}^\mathrm{m}$) components of MOKE to obtain $r_\Phi^\mathrm{m}$, and estimate $r_\theta^\mathrm{s}$ and $r_\eta^\mathrm{s}$ in order to compare them with $r_\mathrm{H}$.

\subsubsection{Experimental verification}
Here we show measurement results from a W/CoFeB bilayer, which was deposited on a different batch of substrates under a slightly different condition compared to the film whose results are presented in the main text.
As the thickness of each layer and the SiO$_2$ are likely different between the two samples, we expect the MOKE signal to differ.
\begin{table*}[htb]
    \caption{Parameters obtained from the measurements and calculations to estimate $r_\theta^\mathrm{s}$ and $r_\eta^\mathrm{s}$.}
    \centering
    \begin{tabular}{cccccccccc}
        \toprule 
           $\theta_\mathrm{P}^\mathrm{m}$ & $\theta_\mathrm{Q}^\mathrm{m}$ & $r_\theta^\mathrm{m}$ & $\eta_\mathrm{P}^\mathrm{m}$ & $\eta_\mathrm{Q}^\mathrm{m}$ & $r_\eta^\mathrm{m}$ & $\theta_\mathrm{P}^\mathrm{s}$ & $\eta_\mathrm{P}^\mathrm{s}$ & $r_\theta^\mathrm{s}$ & $r_\eta^\mathrm{s}$\\
           mrad & mrad &  & mrad & mrad & & mrad & mrad & & \\
        \midrule
        $2.4 \pm 0.04$ & $0.2 \pm 0.4$ & $0.083 \pm 0.2$ & $39 \pm 0.05$ & $0.4 \pm 1.2$ & $0.01 \pm 0.03$ & $-5.3$ & $3.2$ & $0.0078$ & $0.018$\\
        \bottomrule
    \end{tabular}
    \label{table:moke:meas:intrinsic}
\end{table*}

Parameters $\theta_\mathrm{P}$, $\theta_\mathrm{Q}$, $\eta_\mathrm{P}$, $\eta_\mathrm{Q}$ are obtained using the approach described in the main text.
We use the HWP (QWP) for the HWP/QWP in Fig.~\ref{fig:opticalsetup} to measure the real (imaginary) part of $\Phi_\mathrm{K}$.
The obtained values are presented as $\theta_\mathrm{P}^\mathrm{m}$, $\theta_\mathrm{Q}^\mathrm{m}$, $\eta_\mathrm{P}^\mathrm{m}$, $\eta_\mathrm{Q}^\mathrm{m}$ in Table~\ref{table:moke:meas:intrinsic}.
According to Eq.~(\ref{eq:rs}), one must know $\theta_\mathrm{P}^\mathrm{s}$ and $\eta_\mathrm{P}^\mathrm{s}$ in order to estimate $r_\theta^\mathrm{s}$ and $r_\eta^\mathrm{s}$.
Here we assume that $\theta_\mathrm{P}^\mathrm{s}$ and $\eta_\mathrm{P}^\mathrm{s}$ are equivalent to those of a thick CoFeB layer, where optical interference effect can be neglected.
We therefore use data from Ref.~\cite{sumi2018scirep} for $\theta_\mathrm{P}^\mathrm{s}$ and $\eta_\mathrm{P}^\mathrm{s}$ of a thick CoFeB film: see Table~\ref{table:moke:meas:intrinsic} for the extracted numbers. 
Note that $\theta_\mathrm{P}^\mathrm{s}$ and $\eta_\mathrm{P}^\mathrm{s}$ are consistent with the plots presented in Fig.~\ref{fig:sio2dep}(b) as the calculations are based on the parameters described in Ref.~\cite{sumi2018scirep}.
Substituting $r_\Phi^\mathrm{m}$ and $\frac{\eta_\mathrm{P}^\mathrm{s}}{\theta_\mathrm{P}^\mathrm{s}}$ into Eq.~(\ref{eq:rs}), we obtain $r_\theta^\mathrm{s}\sim 0.0078$ and $r_\eta^\mathrm{s} \sim 0.018$: see Table~\ref{table:moke:meas:intrinsic}.
Clearly, $r_\theta^\mathrm{s}$ and $r_\eta^\mathrm{s}$ are significantly smaller than 1 (and $r_\mathrm{H}$).

\begin{acknowledgments}
The authors thank Y. Kobayashi and S. Wang for technical support. This work was partly supported by JSPS KAKENHI (Grant Number 23H00176), JST CREST (JPMJCR19T3), MEXT Initiative to Establish Next-generation Novel Integrated Circuits Centers (X-NICS) and Cooperative Research Project Program of RIEC, Tohoku University.
\end{acknowledgments}

\bibliography{refs_051625.bib}

\end{document}